\documentclass{aa}

\usepackage{amsmath}
\usepackage{graphicx}
\usepackage{array}
\usepackage{booktabs}
\usepackage{threeparttable}
\usepackage{multirow} 
\usepackage{xcolor}
\usepackage{tabularx}
\usepackage{makecell}
\usepackage{amsmath}
\usepackage[amsmath,hyperref]{ntheorem}  

\usepackage{hyperref}
\usepackage{bookmark}  

\begin{document}

   \title{Comparing the Space Densities of Millisecond-Spin Magnetars and Fast X-Ray Transients}
 \author{S. Biswas\inst{1}\thanks{\email{sumedha.biswas@ru.nl}},
          P. G. Jonker\inst{1,2},
          M. Coleman Miller\inst{3},
          A. Levan\inst{1,4},
          J. Quirola-V\'asquez\inst{1}
          } 

   \institute{Department of Astrophysics/IMAPP, Radboud University, PO Box 9010, 6500 GL, The Netherlands
         \and
            SRON, Netherlands Institute for Space Research, Niels Bohrweg 4, Leiden, 2333 CA, The Netherlands
        \and
             Department of Astronomy and Joint Space-Science Institute, University of Maryland, College Park, 20742, Maryland, USA
        \and
            Department of Physics, University of Warwick, Coventry, CV4 7AL, UK
             }

   \date{Received XXX; accepted YYY}

  \abstract
   {Fast X-Ray Transients (FXTs) are bright X-ray flashes with durations of a few minutes to hours, peak isotropic luminosities of $L_{\rm X,peak}\sim 10^{42}-10^{47}$~erg~s$^{-1}$, and total isotropic energies of $E \sim 10^{47}-10^{50}$~erg, which have been detected with space-based telescopes such as \textit{Chandra}, \textit{XMM-Newton}, \textit{Swift-XRT}, and \textit{Einstein Probe} in the soft X-ray band. \textit{Einstein Probe} has detected $>$50 in its first year of operation. While several models have been proposed, the nature of many FXTs is currently unknown. 
   One model predicts that FXTs are powered by the spin-down energy of newly-formed millisecond magnetars. In this context, they are usually thought to form in a binary neutron star (BNS) merger.  However, the rates seem to be in tension: the BNS volumetric rate is estimated to be $\sim 10^2$~Gpc$^{-3}$~yr$^{-1}$, which barely overlaps with the estimated FXT volumetric rate of $10^3-10^4$~Gpc$^{-3}$~yr$^{-1}$, and thus even in the small range of overlap BNS mergers would need to produce FXTs with nearly 100\% efficiency.}
   {We explore the maximum volumetric formation rate of millisecond spin period magnetars, including several possibilities beyond the BNS channel, comparing it with the volumetric rate of FXTs, to determine what fraction of FXTs could arise from a millisecond magnetar origin.}
   {We compile the estimated rate densities for several different suggested formation channels of rapidly spinning magnetars, including the accretion-induced collapse of white dwarfs, binary white dwarf mergers, neutron star -- white dwarf mergers, and the collapse of massive stars. We convert the Milky Way event rates to volumetric rates, wherever necessary, by considering either the star formation rate or the stellar mass density distributions as a function of redshift. 
   }
   {We find that the highest possible rates among these possibilities come from binary white dwarf mergers and the collapse of massive stars. However, both scenarios may be unfavourable for FXT production, due to uncertainties in the resultant spin and magnetic field distributions of the newly formed neutron stars, and several observational constraints. Moreover, in all the scenarios, we find that the fraction of neutron stars that meet both criteria of rapid rotation and a strong magnetic field strength is either very low or highly uncertain. We conclude that millisecond magnetars are not the most viable progenitors of FXTs, and can account for at most 10\% of the entire FXT population. 
   } 
   {}

   \titlerunning{Rate of Formation of Millisecond Magnetars}
   \authorrunning{S. Biswas et al.}
   
   \maketitle
%

\section{Introduction}
Fast X-ray transients (FXTs) are short X-ray flashes with durations of a few minutes to hours, which have been observed in the $\sim$0.3-10 keV X-ray band by \textit{Chandra, XMM-Newton, Swift-XRT} and \textit{Einstein Probe} (e.g., \citealt{soderberg2008, jonker2013, glennie2015, irwin2016, bauer2017, lin2018, xue2019, alplarsson2020, novara2020, lin2020, ide2020, pastormarazuela2020, 2020wilms, lin2021, lin2022, deepak2, quirola1, quirola2, EP240315a, gillanders_2024, liu2024_ep240315}). They are characterized by a single highly energetic outburst, with peak isotropic luminosities of $L_{\rm{X, peak}} \sim 10^{42} - 10^{47} \rm{erg~s^{-1}}$ . So far, $\sim 30$ extragalactic FXTs have been identified through careful archival searches. The first FXT detected with contemporaneous multi-wavelength observations was XRT 080109/SN 2008D, which was serendipitously observed by \textit{Swift} during a supernova observation (\citealt{soderberg2008, mazzali2008, modjaz2009}). 

In early 2024, \textit{Einstein Probe (EP)} was launched \citep{Yuan_2015, Yuan_2022}, and it has announced the discovery of several FXTs in near real-time (e.g., \citealt{EP_FXT_1, EP_FXT_2, EP_FXT_3, EP_FXT_4, EP_FXT_5, EP_FXT_6}). This has led to prompt multi-wavelength follow-up observations of FXTs, helping us understand their origin. 
For example, \cite{EP_FXT_2} reported EP~240315a and it led to the first detection of both an optical and radio counterpart of an EP-discovered FXT, detected at a redshift of $z$ = 4.859 \citep{gillanders_2024, 2024GCN_srivastav, 2024GCN_saccardi, 2024GCN_carotenuto, 2024GCN_bruni, 2024GCN_leung, EP240315a}. \cite{EP240315a} concluded that the observed properties of EP~240315a are potentially consistent with a long GRB (collapsar) interpretation of FXTs (also see \citealt{gillanders_2024} and \citealt{liu2024_ep240315}). 

While the origin of the majority of FXTs is unknown, several theories have been put forward: (1) X-ray emission produced during the spindown of a millisecond magnetar produced in a binary neutron star (BNS) merger (e.g., \citealt{dai2018, fong2015, sun2017, bauer2017, xue2019}); (2) supernova shock breakouts (SBO) from core-collapse supernovae (CCSNe), where the X-ray emission is generated from the breakout of the supernova explosion once it crosses the surface of an evolved star such as a blue supergiant or a Wolf-Rayet star (e.g., \citealt{soderberg2008, nakarsari2010, waxmankatz2017, novara2020, alplarsson2020}); (3) tidal disruption event (TDE) involving a white dwarf (WD) and intermediate mass black hole (IMBH), where X-rays are produced by the  accretion disc and/or relativistic jet  (e.g., \citealt{jonker2013, macleod2014, saxton2021, maguire2020}); (4) X-ray emission from off-axis or sub-luminous gamma-ray bursts (GRBs), produced by the mildly relativistic cocoon jet once it breaks the surface of a massive progenitor star (e.g., \citealt{ramirezruiz2002, zhang2004, nakar2015, zhang2018, delia2018}). Several of these theories have been probed by studying the host galaxies of some FXTs, and the offset of the FXT with respect to the centre of the host, where a larger offset potentially implies an older system (e.g., \citealt{deepak1, deepak2, deepak2024, inkenhaag2024redshiftscandidatehostgalaxies, 2025quirola}).

In this work, we investigate the millisecond magnetar model for FXTs.  The ultimate source of energy for millisecond magnetar-powered FXTs is rotational, with $E_{\rm rot}\approx 2\times 10^{52}~{\rm erg}~(I/10^{45}~{\rm g~cm}^2)(\nu/1000~{\rm Hz})^2$, where $I$ and $\nu$ are the moment of inertia and the spin frequency of the neutron star, respectively.  Thus, there is in principle plenty of rotational energy in the millisecond magnetars to explain FXTs. If the duration of an FXT is similar to its magnetic dipole spindown time, then a $B\sim 10^{14}-10^{16}$~G with a birth frequency $\nu=1000$~Hz would be consistent with the typically observed duration of a few minutes to a few hours for individual FXTs and luminosities of $L_{X, peak} \sim 10^{42} - 10^{47} \rm{erg~s^{-1}}$, assuming that the efficiency $\eta$ of X-ray production is of the order of $\sim 10^{-3}-10^{-2}$ ($\eta$ is defined as $L_{\rm X}/L_{\rm sd}$, where $L_{\rm X}$ is the X-ray luminosity and $L_{\rm sd}$ is the spin-down luminosity).

If millisecond magnetars originate from binary neutron star (BNS) mergers, then a positive feature of this model is that these mergers typically expel only $\sim 10^{-4} - 10^{-2}~\rm{M_\odot}$ (e.g., \citealt{2013hotokezaka}) of material. While this ejecta contains significant amounts of r-process elements with high opacity, its rapid expansion leads to decreasing density over time, potentially allowing X-ray radiation to escape as the ejecta becomes optically thin. Previous studies have modelled and calculated the optical/UV and X-ray emission expected from the spindown of millisecond magnetars produced from BNS mergers, and predicted their light curves (e.g., \citealt{zhang2013, siegelciolfi_1, siegelciolfi_2, sun2017}). Two FXTs were identified in the 7 Ms \textit{Chandra} Deep Field-South (CDF-S) data set, XRT~141001 \citep{bauer2017} and XRT~150322 \citep{xt2}, denoted also as "CDS-XT1" and "CDS-XT2", respectively. Motivated by the above spin-down magnetar models, \cite{sun2019} (see their figure~2) interpret both CDS-XT1 and CDS-XT2 within the framework of the BNS merger millisecond magnetar model. 

However, the rate of BNS mergers is in tension with the rate of FXTs. The volumetric rate of known distant extragalactic FXTs is $\sim~10^3 - 10^4~\rm{{Gpc}^{-3}~{yr}^{-1}}$  \citep{quirola1, quirola2}. In comparison, the volumetric rate of BNS mergers is $\sim~10^1 - 10^3~\rm{{Gpc}^{-3}~{yr}^{-1}}$ from GW detections through LVK run O3 (e.g., \citealt{Mandel_2022, gwtc3}). Although the upper limit of the BNS merger rate overlaps with the lower limit of the FXT rate, it would require an efficiency close to unity to account for all FXTs, i.e., assuming that each BNS merger results in an FXT. Depending on the neutron star equation of state, the masses of the two neutron stars before merger, and their mass ratio, the merger product could however, also be a black hole \citep{Metzger2019}. 

Here we therefore also consider the formation of millisecond magnetars through other channels such as the accretion-induced collapse (AIC) of white dwarfs (e.g., \citealt{1980miyaji}), binary white dwarf (BWD) mergers (e.g., \citealt{levan2006}), neutron star - white dwarf (NSWD) mergers (e.g., \citealt{Metzger_NSWD}), and the collapse of massive stars (e.g., \citealt{Beniamini_2019}). 

In Section~\ref{formscen}, we review the formation scenarios of rapidly spinning magnetars. In Section~\ref{combinedrate}, we discuss the combined rate of formation of magnetars (Table~\ref{tab:rates}), and we discuss other constraints such as the uncertainties in the spin distributions, the effects of the surrounding post-merger ejecta, the duration of FXTs, and the consequent implications for their origin. Wherever relevant, we have converted the event rates to volumetric rates using one of the two methods described in Appendix~\ref{rateconversion}. We conclude in Section~\ref{conclusion}.

\section{Rate of Formation of Rapidly Spinning Magnetars} \label{formscen}

\begin{figure*}
    \centering
    \includegraphics[width=1\linewidth]{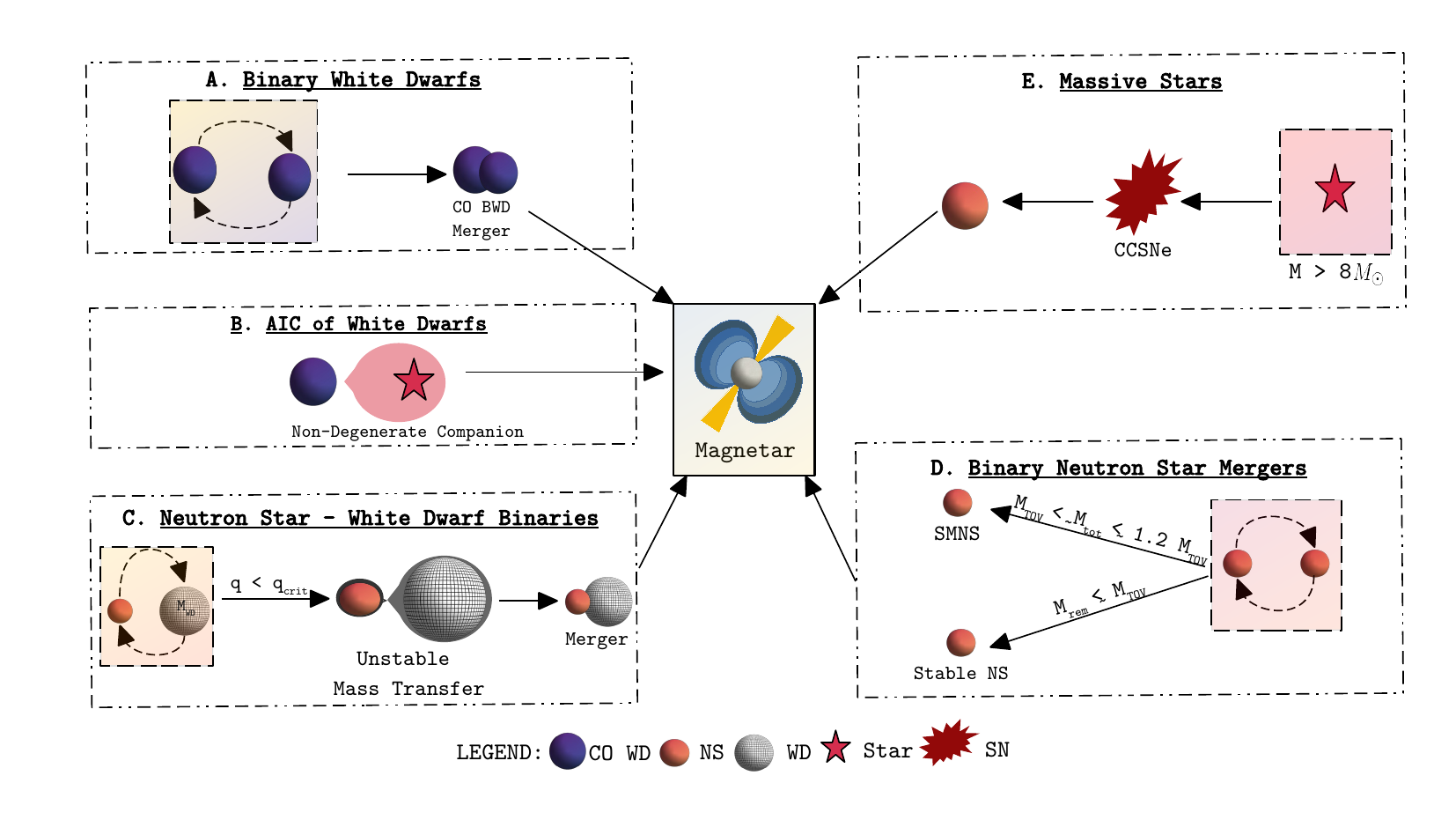}
    \caption{Schematic diagram to summarize the different formation mechanisms of rapidly spinning magnetars: (A) Binary white dwarf mergers (B) Accretion-induced collapse of massive white dwarfs (C) Neutron star-white dwarf mergers (D) Binary neutron star mergers (E) Collapse of massive stars}
    \label{fig:form-mech}
\end{figure*}

In the following sections, we describe the various formation pathways for rapidly spinning magnetars as summarized in Figure~\ref{fig:form-mech}. 

\subsection{\textbf{Pathway A:} Binary White Dwarfs}

The Milky Way (MW) contains $\sim 10^{10}$ WDs, of which about $(2 - 3) \times 10^8$ are thought to exist in binaries with other WDs \citep{nelemans2001, 2009holberg, 2009Napiwotzki}. About half of these binaries are predicted to merge within a Hubble time. The BWD merger rate has been estimated by, for example, \cite{nelemans2001} to be $3 \times 10^{-3}~\rm{yr^{-1}}$ in the MW for binaries with total masses higher than $\rm M_{Ch}$, where $\rm M_{Ch} \approx 1.4~\rm{M_\odot}$ is the Chandrasekhar mass \citep{1931chandrasekhar}.

The final outcome of a BWD merger, whether it produces a magnetar or not, depends on the properties of the WDs involved. Different combinations of WD binaries, for example, CO+CO, CO+ONe, and ONe+ONe\footnote{CO: Carbon-Oxygen; ONe: Oxygen-Neon;}) lead to different merger remnants (e.g., \citealt{dan2014, 2020liu}). However, their outcomes are not well understood yet. Here, we consider CO+CO WD binaries, which account  for $\sim$25\% of all BWDs in the MW \citep{nelemans2001}, although some recent studies also suggest other types of WD mergers such as ONe+CO WD binaries as progenitors for NSs (and potentially magnetars) (e.g., \citealt{Kashyap_2018, ONe_wu2023}). We consider two CO WDs in a binary, with a total binary mass of $\rm M = M_{WD1} + M_{WD2}$, where $\rm M_{WD1}$ and $\rm M_{WD2}$ are the individual masses of the WDs that eventually merge. 

If $\rm M > M_{Ch}$, the CO WD merger would typically result in the explosion of the more massive WD (e.g., \citealt{1996hachisu, 1997li_vandenheuvel, 2009wang, 2012shen, 2013wang, 2016wu, 2018wang, 2019perets}), resulting in a Type Ia supernova (SN). However, if one or both WDs have very strong magnetic field strengths ($\gtrsim 2 \times 10^6 \rm{G}$), it could produce a rapidly spinning magnetar after the merger \citep{2001king, levan2006}, alternatively, angular momentum may be transferred from the center to the outer regions, enabling the formation of a disk around the newly formed compact object  (e.g., \citealt{K_lebi_2013}) About 10\% of the WD population has sufficiently strong magnetic field strengths, which implies an upper limit for the rate of formation of a msec.~spin magnetar as $3 \times 10^{-4}~\rm{yr^{-1}}$ in the MW \citep{levan2006}.  

\subsection{\textbf{Pathway B:} Accretion-Induced Collapse of Massive White Dwarfs}
Another scenario for the formation of a NS (and potentially a millisecond magnetar) is the accretion-induced collapse (AIC) of a massive WD, after it accretes matter from a non-degenerate companion such as a main-sequence star, a red giant, or a He star (e.g., \citealt{tauris2013, schwab2016}), and its mass increases. If $M > M_{Ch}$, the massive WDs are predicted to collapse into NSs due to the rapid electron-captures onto heavy elements produced by the oxygen burning, preventing a thermonuclear explosion (e.g., \citealt{1980miyaji}). NS formation is considered to be the most likely outcome of AIC of a WD, however, also see \citet{2016jones, 2019jones} for discussions about alternative scenarios where a thermonuclear explosion occurs. The produced NS has millisecond spins due to the conservation of angular momentum; as the WD collapses, its radius decreases, causing its rotation rate to increase significantly \citep{1992duncan, 2006pricerosswog, 2013zrake_macfayden}. Additionally, the rapid rotation, combined with the dynamics of the collapse, may lead to the amplification of any pre-existing magnetic fields and the generation of strong magnetic fields, resulting in surface dipole fields on the order of $\rm 10^{14} - 10^{15}$~G, characteristic of magnetars \citep{Piro_Koll_2016}. \cite{1998yungelson} determined the AIC event rate in the MW to be $\rm 8 \times 10^{-7} - 8 \times 10^{-5}~yr^{-1}$, and \cite{Piro_Koll_2016} further estimated that a fraction of 0.01 -- 0.1 of AIC of massive WD events produce a millisecond magnetar, corresponding to $\rm \sim 10^{-9} - 10^{-6}~yr^{-1}$. 

\subsection{\textbf{Pathway C:} Neutron Star-White Dwarf Binaries}

A standard scenario for the formation of a tight NS-WD binary invokes common envelope evolution of an initially wide binary, consisting of a NS and an intermediate-mass ($\lesssim$8--10$~\rm{M_\odot}$) main-sequence companion (e.g. \citealt{vandenheuvel_1984}) which eventually forms a WD.  The composition of the WD has a high probability of being CO, but it could also be ONe, Helium (He) or He-CO \citep{toonen2018}. Eventually, a tight NS-WD binary is formed, and the system continues to lose orbital angular momentum on long timescales due to gravitational wave emission (see \citealt{1964peters} for inspiral time calculations). 

The outcome of the merger of two compact objects with masses $\rm M_{WD}$ and $\rm M_{NS}$ in a NS-WD binary is dependent upon the critical mass ratio $\rm q_{crit} = M_{WD, crit}/M_{NS}$, where $\rm M_{WD, crit}$ is the critical WD mass. Bobrick et al. (2017) established $\rm M_{WD, crit} = 0.2~M_\odot$, a value lower than previous estimates but considered to be more robust due to its incorporation of efficient angular momentum loss through disc winds. \cite{toonen2018}  suggested that over 99.9\% of semidetached NS–WD binaries would merge when $\rm M_{WD, crit} = 0.2~M_\odot$. There are $\gtrsim$ 20 NS-WD binaries identified in the MW \citep{lorimer2005}, of which 4 are predicted to merge in $\lesssim 10^{10}$ years (see \citealt{Metzger_NSWD} for references). Close NS-WD binaries can also be formed directly by collisions in dense stellar regions, such as the centres of galaxies or globular clusters \citep{1997sigreed}.

Based on the observed properties of the NS-WD binaries in our MW (such as their masses and orbital periods), \cite{2004kim} estimated a merger rate of $10^{-6} - 10^{-5}~\rm{yr^{-1}}$ in the MW. Some population synthesis models predict higher rates of $10^{-5} - 10^{-3}~\rm{yr^{-1}}$ in the MW (e.g., \citealt{1999_portegies, 2000tauris, 2002davies}). \cite{2009thompson} estimated the volumetric event rate of NS-WD mergers in the local universe to be $(0.5 -1) \times 10^4~\rm{Gpc^{-3}~yr^{-1}}$.

If $\rm q < q_{crit}$, unstable mass transfer takes place as the WD is tidally disrupted by the NS on dynamical timescales, leading to a merger (e.g., \citealt{1987hjellming_webbink}). \cite{margalitmetzger2016} suggested that a NS-WD merger would lead to the formation of a millisecond pulsar surrounded by an accretion disk. The magnetic field strength of the NS may either increase via a dynamo winding-up process \citep{pascha2011} or decrease through enhanced ohmic dissipation of accreted matter in the crust of the NS \citep{1997konar, 1997urpin, 2004cumming}. \cite{Zhong_2020} found that during the unstable mass transfer process, the magnetic field of the NS undergoes amplification via an $\alpha-\omega$ dynamo mechanism. This occurs as the accreting NS becomes enveloped by a massive, extended hot disk composed of WD debris, potentially resulting in the formation of a millisecond magnetar. It is difficult to ascertain the exact fraction of magnetars formed from this NS-WD merger pathway. If we follow the assumption made in \cite{Zhong_2020} that the magnetar formation rate through this channel is similar to the BNS merger rate i.e., 3\% of all NS-WD mergers \citep{Nicholl_2017}, then the volumetric rate is $150-300~\rm{Gpc^{-3}~yr^{-1}}$. However, as there is no clear justification given for this assumption, and as this scenario requires further numerical/observational studies, we note that the millisecond magnetar formation rate through this channel is highly uncertain.

\subsection{\textbf{Pathway D:} Binary Neutron Stars}
The merger of two NSs has four possible outcomes (see table~3 in \citealt{Metzger2019} for mass ranges, lifetimes and rates of each outcome) that for instance, depend on the total mass of the binary, $\rm M_{tot}$ \citep{shibata_uryu_2000, 2006_shibata_taniguchi}, and the nuclear equation-of-state (EoS) which sets the maximum allowed non-rotating NS mass, $\rm M_{TOV}$. For example, \cite{2017margalitmetzger} determine $\rm M_{TOV} \lesssim 2.7~M_\odot$ (see their figure 4 and table 2 for details about the dependence on the EoS). 

If $\rm M_{rem} \lesssim M_{TOV}$, the merger results in a stable NS that can live indefinitely. For $\rm M_{TOV} \lesssim M_{tot} \lesssim 1.2,M_{TOV}$, a supramassive NS (SMNS) is produced, which remains stable for $\gg 300$ ms before collapsing to a black hole (BH). When $\rm M_{crit} \gtrsim M_{tot} \gtrsim M_{TOV}$, a hypermassive NS (HMNS) forms with strong differential rotation and typically collapses into a BH within seconds to minutes \citep{Metzger2019}. If $\rm M_{tot} > M_{crit}$, where $\rm M_{crit} \sim 2.6 - 3.9~M_\odot$ \citep{2011hotokezaka, 2013bauswein}, the remnant collapses to form a BH almost immediately after merger. If the remnant NS survives the merger, it is expected to have a millisecond spin period due to conservation of orbital angular momentum during the inspiral and merger phases (e.g., \citealt{shibata_uryu_2000, 2011hotokezaka}). Furthermore, during the merger, shear flows and turbulence driven by the Kelvin-Helmholtz instability enhance magnetic field strengths (B~$\sim 10^{14} - 10^{16}$~G) through small-scale dynamo mechanisms (e.g., \citealt{2015giaco}).

However, remnant stability is not guaranteed by the condition of $\rm  M < 1.2~M_{TOV}$ alone; \cite{2021beniamini} show that SMNSs collapse into a BH after shedding $\sim (3 -6)\times10^{52}~\mathrm{erg}$ of rotational energy before collapsing into a BH, with survival timescales sensitive to their magnetic field strength and angular momentum transport efficiency. Furthermore, \cite{2022margalit} demonstrate that the angular momentum profile of the SMNS influences its stability: differential rotation provides temporary centrifugal support, but magnetorotational instabilities (MRI) redistribute angular momentum in regions of decreasing angular velocity, while solid-body rotation in the core enhances stability. These effects imply that only a fraction of SMNS remnants ($\rm M_{tot} < 1.2M_{TOV}$) avoid rapid collapse, depending on the energy loss and angular momentum redistribution timescales.

The discovery of NSs with masses $\sim 2~\rm{M_\odot}$ \citep{2010demorest, 2013antoniadis, 2022romani} establishes a lower limit on $\rm M_{TOV}$, while the upper limit of $\sim 2.1 - 2.2~\rm{M_\odot}$ can be inferred from the detection of GW170817 and the corresponding EM counterpart detections; these estimates depend on assumptions such as the EoS (e.g., \citealt{2017margalitmetzger, 2017granot, 2017shibata, Shao_2020}). While BNS mergers with total mass $\rm M_{tot} > M_{crit}$ result in prompt BH formation, the existence of massive NSs implies that some merger remnants must have masses $\rm < M_{crit}$, allowing for the formation of millisecond magnetars rather than promptly collapsing into BHs.

The rate of formation of millisecond magnetars could potentially be constrained observationally in several ways, assuming current theoretical models of kilonovae are accurate. First, a millisecond magnetar central engine is predicted to release more energy than is observed in SGRBs, as indicated by early-time X-ray and gamma-ray detections-- such as the absence of extended emission or internal plateaus, and late-time radio follow-up observations \citep{2016horesh, 2020schroder, 2021beniamini, 2021ricci}. Secondly, such mergers are predicted to produce kilonovae that are significantly brighter than those without a millisecond magnetar remnant, which are inconsistent with limits from all-sky optical surveys such as ZTF \citep{wang2023constraining}. Lastly, these mergers are expected to produce kilonova afterglows that are brighter and peak earlier than those without a millisecond magnetar remnant, but no such afterglows have been observed till date \citep{2025acharya}. Additionally, the rate of BNS mergers from GW observations is $\sim 10^1 - 10^3~\rm{Gpc^{-3}~yr^{-1}}$ (e.g. \citealt{Mandel_2022, gwtc3}), which provides an upper limit for the rate of formation of millisecond magnetars from this channel. 

\subsection{\textbf{Pathway E:} Massive Stars} \label{massivestars}
Massive stars ($\gtrsim 8~\rm{M_\odot}$) end their lives with a core-collapse supernova (CCSN), leaving behind a compact object (NS or BH), at a rate of $10^5 - 10^6~\rm{Gpc^{-3}~yr^{-1}}$ \citep{2019eldridge}, the majority of these systems are likely to form NSs. It is likely that the majority of the $\sim 30$ magnetars observed within the MW \citep{magnetarcat_mcgill} arise via this channel given their locations in the Galactic plane, associations with star-forming regions (e.g., \citealt{2012tendulkar}), and in some cases, direct locations at the centre of SNe remnants \citep{2022lyman}. Recent estimates suggest that a substantial fraction of neutron stars, $0.4_{-0.28}^{+0.6}$, may be born as magnetars \citep{Beniamini_2019}. This is consistent with previous studies that proposed approximately 10\% of CCSNe form magnetars \citep{1998kouv,2007gill_heyl}, which corresponds to $10^4 - 10^5~ \rm{Gpc^{-3}~yr^{-1}}$, and implies that the magnetar production rate is a substantial fraction of the CCSNe rate. 

However, the fraction of magnetars that are created with the necessary millisecond spin periods to power magnetar-driven transients is much smaller. Indeed, should all magnetars be created with such periods we should observe their enhanced energy input in the light curves of a large fraction of SNe. Instead, observations of SNe suggest that only a small fraction is consistent with the high luminosities that would be obtained via magnetar spin-down power (e.g., \citealt{Rea_2015}, also see \citealt{2014nakar} and \citealt{2024rodriguez}). Indeed, \cite{Rea_2015} suggest that extreme transients require an additional population of "super-magnetars" (millisecond magnetars), substantially rarer than the general population. These objects are estimated to form at a rate of $\rm \leq 16~Myr^{-1}$ in the MW -- significantly rarer than the CCSNe event rate in the MW $\rm \sim 10^4~Myr^{-1}$ \citep{Rozwadowska_2021}. This implies that $<0.2\%$ of CCSNe events could potentially produce millisecond magnetars, i.e., $\rm \sim 20~Myr^{-1}$ in the MW or $\rm < 10^2~Gpc^{-3}~yr^{-1}$. 

The challenge in obtaining millisecond magnetars in core collapse is that angular momentum conservation must enable a nascent NS with a rotation period of millisecond duration. Specifically, millisecond magnetars, rather like BH engines in GRBs require specific angular momentum (angular momentum per unit mass) of $\rm j \gtrsim 10^{16}$ cm$^{2}$ s$^{-1}$ \citep{2016levan}. However, massive single stars spin down due to mass and angular momentum loss via stellar winds and obtaining the necessary rotation requires binary interactions (e.g., \citealt{Ghodla_2022}), for example. While substantial uncertainties remain on how the core couples to its envelope, and in the detailed stellar evolution it is clear that only a small minority of core collapse events can create magnetars with millisecond spin periods, and that this fraction may be strongly metallicity dependent.

\section{Discussion} \label{combinedrate}

\subsection{Event Rate Conversion} \label{rateconversion}
Throughout this paper, we refer to literature that provides the event rates in different units. Several papers estimate event rates in the context of the Milky Way (MW) (e.g., \citealt{nelemans2001, levan2006}) while some others calculate the volumetric event rates (e.g., \citealt{gwtc3, Zhong_2020}). To be able to compare the different formation scenarios and the FXT event rate, it is important to convert all the rates to the same volumetric units of $\rm{Gpc^{-3}~yr^{-1}}$. We describe the method we use for this below. 

To calculate the volumetric rate $\mathcal{R}$, we follow the form of equation~7 in \citealt{2008kopparapu} and rewrite it considering two different cases: $\mathcal{R}_{\rm SMD}$ and $\mathcal{R}_{\rm SFR}$. First, we consider scenarios that depend on the galaxy stellar mass density (SMD) present at each epoch, rather than ongoing star formation. This is more suitable for binary merger scenarios, especially those with longer delay times between star formation and the merger. The galaxy SMD decreases with increasing redshift, reflecting the buildup of stellar mass as the universe ages. We have, 
\begin{equation}
    \mathcal{R}_{\rm SMD} = \frac{r_{\rm{MW}}}{\rm{M}_{\rm{MW}}} \rho^*,
\end{equation}
where $r_{\rm{MW}}~\rm{[yr^{-1}]}$ is the event rate in the MW, $M_{\rm{MW}}\sim 10^{10}~\rm{M_\odot}$ (e.g., \citealt{Licquia_2015}) is the stellar mass of the MW, and $\rho^*~\rm{[M_\odot~Gpc^{-3}]}$ is the galaxy SMD (see table C.1. in \citealt{2023weaver}). 

Second, we consider the volumetric rate as a function of the evolution of the star formation rate (SFR) with redshift, this is relevant for formation mechanisms that are strongly linked to star formation, such as the collapse of massive stars and binary mergers with short delay times. Thus, we have 
\begin{equation}
    \mathcal{R}_{\rm SFR} = \frac{r_{\rm{MW}}}{\rm{SFR}_{\rm{MW}}} \rm{SFR}(z),
\end{equation}
where $\rm{SFR}_{\rm{MW}}\sim 2.0 \pm 0.7~\rm{M_\odot}~yr^{-1}$ is the current star formation rate of the MW (e.g., \citealt{Elia_2022}), and SFR(z) is the SFR density as a function of redshift, given by \cite{2014madau_dickinson}:
    \begin{equation} \label{sfr}
        \rm{SFR}(z) = 0.015 \frac{(1+z)^{2.7}}{1 + [(1+z)/2.9]^{5.6}}~\rm{M}_{\odot}~\rm{Mpc}^{-3}~ \rm{yr}^{-1}.
    \end{equation}
The SFR(z)  distribution peaks at $z \sim$ 2, so the volumetric rate is highest at that redshift. 

\begin{figure*}
    \centering
    \includegraphics[width=\linewidth]{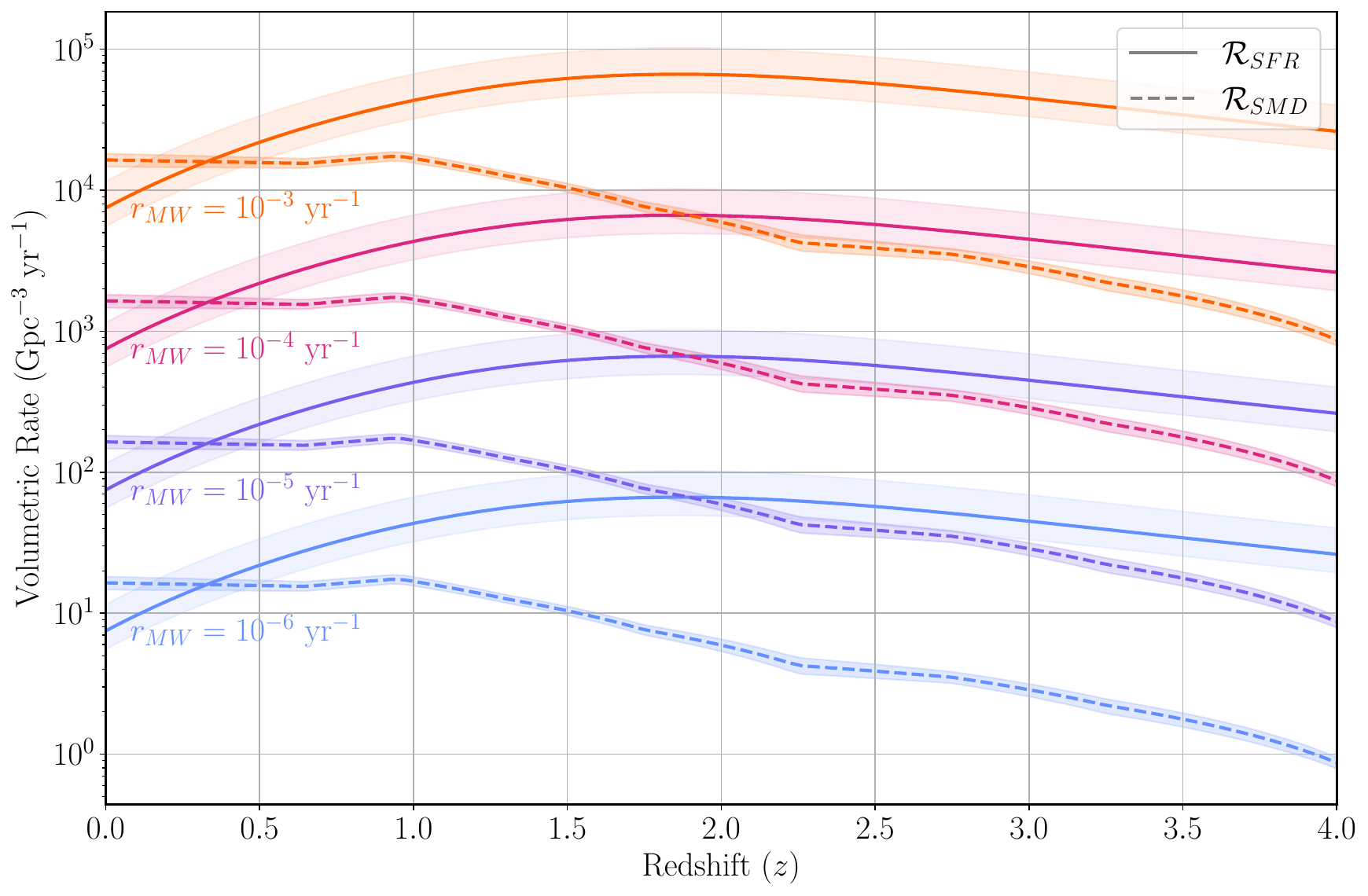}
    \caption{Volumetric rates as a function of redshift calculated using two different methods: galaxy SMD ($\rm \mathcal{R}_{SMD}$ dashed lines) suitable for binary mergers with longer delay times between star formation and the merger, and SFR density ($\rm \mathcal{R}_{SFR}$, solid lines) suitable for events stronly linked to star formation such as the collapse of massive stars and binary mergers with shorter delay times between star formation and the merger. The rates are computed for different MW event rates ($\rm r_{MW}$) ranging from $10^{-6}$ to $10^{-3}$ $\rm yr^{-1}$.}
    \label{fig:sfr_smd_z}
\end{figure*}

\begin{table*}
  \centering
  \caption{Volumetric rates calculated using SFR density ($\rm \mathcal{R}_{SFR}$) and galaxy SMD ($\rm \mathcal{R}_{SMD}$) methods at different redshifts for various MW event rates $r_{\rm MW} [\rm{yr}^{-1}]$}.
  \label{tab:vol_rates}
  \begin{threeparttable}
    \begin{tabular}{c|cccc}
      \toprule
      \textbf{$z$} & \boldmath$r_{\rm MW}=10^{-3}$ & \boldmath$r_{\rm MW}=10^{-4}$ & \boldmath$r_{\rm MW}=10^{-5}$ & \boldmath$r_{\rm MW}=10^{-6}$ \\
      \midrule
      \multicolumn{5}{l}{\textbf{$\rm \mathcal{R}_{SFR}$} (Gpc$^{-3}$ yr$^{-1}$)} \\[1ex]
      $z = 0.3$ & $(1.71^{+0.92}_{-0.44}) \times 10^4$ & $(1.71^{+0.92}_{-0.44}) \times 10^3$ & $(1.71^{+0.92}_{-0.44}) \times 10^2$ & $(1.71^{+0.92}_{-0.44}) \times 10^1$ \\[1ex]
      $z = 1.0$ & $(4.38^{+2.36}_{-1.14}) \times 10^4$ & $(4.38^{+2.36}_{-1.14}) \times 10^3$ & $(4.38^{+2.36}_{-1.14}) \times 10^2$ & $(4.38^{+2.36}_{-1.14}) \times 10^1$ \\[1ex]
      $z = 2.0$ & $(6.58^{+3.54}_{-1.70}) \times 10^4$ & $(6.58^{+3.54}_{-1.70}) \times 10^3$ & $(6.58^{+3.54}_{-1.70}) \times 10^2$ & $(6.58^{+3.54}_{-1.70}) \times 10^1$ \\[1ex]
      $z = 3.0$ & $(4.42^{+2.38}_{-1.15}) \times 10^4$ & $(4.42^{+2.38}_{-1.15}) \times 10^3$ & $(4.42^{+2.38}_{-1.15}) \times 10^2$ & $(4.42^{+2.38}_{-1.15}) \times 10^1$ \\[1ex]
      $z = 4.0$ & $(2.61^{+1.41}_{-0.68}) \times 10^4$ & $(2.61^{+1.41}_{-0.68}) \times 10^3$ & $(2.61^{+1.41}_{-0.68}) \times 10^2$ & $(2.61^{+1.41}_{-0.68}) \times 10^1$ \\[2ex]
      \multicolumn{5}{l}{\textbf{$\rm \mathcal{R}_{SMD}$} (Gpc$^{-3}$ yr$^{-1}$)} \\[1ex]
      $z = 0.3$ & $(1.59^{+0.15}_{-0.14}) \times 10^4$ & $(1.59^{+0.15}_{-0.14}) \times 10^3$ & $(1.59^{+0.15}_{-0.14}) \times 10^2$ & $(1.59^{+0.15}_{-0.14}) \times 10^1$ \\[1ex]
      $z = 1.0$ & $(1.67^{+0.13}_{-0.12}) \times 10^4$ & $(1.67^{+0.13}_{-0.12}) \times 10^3$ & $(1.67^{+0.13}_{-0.12}) \times 10^2$ & $(1.67^{+0.13}_{-0.12}) \times 10^1$ \\[1ex]
      $z = 2.0$ & $(5.80^{+0.60}_{-0.58}) \times 10^3$ & $(5.80^{+0.60}_{-0.58}) \times 10^2$ & $(5.80^{+0.60}_{-0.58}) \times 10^1$ & $(5.80^{+0.60}_{-0.58}) \times 10^0$ \\[1ex]
      $z = 3.0$ & $(2.79^{+0.28}_{-0.30}) \times 10^3$ & $(2.79^{+0.28}_{-0.30}) \times 10^2$ & $(2.79^{+0.28}_{-0.30}) \times 10^1$ & $(2.79^{+0.28}_{-0.30}) \times 10^0$ \\[1ex]
      $z = 4.0$ & $(8.70 \pm 0.80) \times 10^2$ & $(8.70 \pm 0.80) \times 10^1$ & $(8.70 \pm 0.80) \times 10^0$ & $(8.70 \pm 0.80) \times 10^{-1}$ \\
      \bottomrule
    \end{tabular}
    \begin{tablenotes}
      \small
      \item $\rm \mathcal{R}_{SFR}$ values are derived assuming SFR$_{\rm MW} = 2.0 \pm 0.7$ M$_\odot$ yr$^{-1}$.
      \item $\rm \mathcal{R}_{SMD}$ values are calculated using $\rho^*$ values from table C.1 in \citealt{2023weaver}.
    \end{tablenotes}
  \end{threeparttable}
\end{table*}

The BWD merger rate in the MW is $\rm 3 \times 10^{-3}~yr^{-1}$ \citep{nelemans2001}, this can be converted to $\rm \mathcal{R}_{SFR} \sim 10^4~Gpc^{-3}~yr^{-1}$ and $\rm \mathcal{R}_{SMD} \sim 10^3 - 10^4~Gpc^{-3}~yr^{-1}$. Similarly, the rate of production of magnetars from BWD mergers is $\rm 3 \times 10^{-4}~yr^{-1}$ \cite{levan2006}, and the NSWD merger rate in the MW is $\rm 10^{-6} - 10^{-3}~yr^{-1}$ \citep{2004kim, 1999_portegies, 2000tauris, 2002davies}; the corresponding converted volumetric rate values for each of these scenarios is listed in Table~\ref{tab:vol_rates} (also see Figure~\ref{fig:sfr_smd_z}). 
progenitor
It is important to note that in our event rate conversion, we consider the stellar masses of the galaxies instead of their blue-light luminosities. Previously, the number of Milky Way Equivalent Galaxies (MWEG), $\rm n_{MWEG}$, was calculated by considering the blue-light luminosity of galaxies, corrected for dust extinction and reddening \citep{1991phinney}, as a measure of star formation. However, blue-light luminosity may not be a perfect tracer of current SFR (e.g., \citealt{2010gw}), and instead Equation~\ref{sfr} provides a better estimate \citep{2014madau_dickinson}. While the MWEG method using blue-light luminosity has been effective for local universe estimates, considering the SFR and SMD densities encompasses redshift evolution. The SFR density peaks at z$\sim$2 with values $\sim$10 times higher than at z$\sim$0, while the SMD shows a continuous increase towards z = 0, making this approach useful for calculating rates of events with different delay times relative to star formation.

\begin{table}
  \centering
  \caption{Volumetric rate of formation of magnetars from compact object scenarios}
  \label{tab:rates}
  \begin{threeparttable}
    \begin{tabularx}{\linewidth}{cccc}
      \toprule
      \textbf{Pathway} & \textbf{Event} & \textbf{\makecell{Magnetar\\Formation\\Rate\tnote{(a)}}} & \textbf{\makecell{Millisecond \\Magnetar\\Formation\\Rate\tnote{(a)}}} \\
      \midrule
      A & BWD & $< 10^3 - 10^5$\tnote{(b)} & $< 10^4$ \\ 
      B & AIC of WD & $< 10^1 - 10^3$\tnote{(b)} & $< 10$ \\
      C & NSWD & 150 -- 300 & -- \\
      D & BNS & $10^1 - 10^3$ & $\lesssim 8.5 \times 10^2$\tnote{(c)} \\
      E & Massive Stars & $10^4 - 10^5$ & $<10^2$ \\
      \bottomrule
    \end{tabularx}
    \begin{tablenotes}
      \small
      \item[(a)] Rates are given in $\rm{Gpc^{-3}~yr^{-1}}$;
      \item[(b)] Event rate, therefore an upper limit for magnetar formation;
      \item[(c)] Estimated from table 3 in \cite{Metzger2019};
    \end{tablenotes}
  \end{threeparttable}
\end{table}

\subsection{Spin Distribution \& Magnetic Field Strength Distribution Uncertainties}

There are significant uncertainties in the initial spin periods and magnetic field strength distribution of newly-formed magnetars in all the formation pathways that we discuss in Section~\ref{formscen}. In this subsection, we discuss those originating from the collapse of massive stars and BWD mergers, as these pathways exhibit the highest volumetric rates for millisecond magnetar production, thereby representing the most viable formation channels. For magnetars formed during the collapse of a massive star, several factors such as the mass of the star, metallicity \citep{Song_2023}, and magnetic field configuration (e.g., \citealt{1992duncan}) contribute to the uncertainty in the initial spin period distribution. Some models suggest millisecond periods for dynamo-generated fields \citep{1992duncan}. \cite{Jawor_2021} uses population synthesis models to provide an upper bound, concluding that the initial spin period of magnetars must be less than 2~s. This is in agreement with observations of SN remnants around some magnetars that suggest initial spin periods $> 5~\rm{ms}$ \citep{2006vink_kuiper, zhou2019}.  

In the case of BWD mergers that produce a magnetar remnant, and assuming conservation of angular momentum, we can link properties of the newly formed magnetar to those of the progenitor WDs by considering the following simplified equations \citep{kremer2023}:

\begin{equation}
    \begin{aligned} & B_{\mathrm{magnetar}}=B_{\mathrm{WD}}\left(\frac{R_{\mathrm{WD}}}{R_{\mathrm{magnetar}}}\right)^{2} \\ & P_{\mathrm{magnetar}}=P_{\mathrm{WD}}\left(\frac{R_{\mathrm{magnetar}}}{R_{\mathrm{WD}}}\right)^{2},\end{aligned}
\end{equation}
where $\rm R_{magnetar}$ = 10~km, and $\rm R_{WD}$ = $10^4$~km. To achieve $\rm B_{magnetar}$ = $10^{14} - 10^{16}$~G and $\rm P_{magnetar}$ = 1 -- 2~ms, we require at least one of the WD progenitors to have $\rm B_{WD} \gtrsim 10^8$~G and $\rm P_{WD} \approx 10^3$~s. 
About 10\% of WDs have fields in excess of $10^6$~G \citep{2007kawka}, with the fraction of WDs with fields $\gtrsim 10^8$~G considerably smaller, i.e., $\sim$2 --  4\% (see figure 2 in \citealt{wickferr2005} and figure 5 in \citealt{2024mohapatra}). Furthermore, observations indicate that magnetic WDs exhibit a wide range of rotational periods, from minutes to years, with the most strongly magnetized WDs tending to be slower rotators \citep{2015ferrario}. These observational constraints suggest that only a very small fraction of WDs meet the criteria necessary for millisecond magnetar formation through BWD mergers. 

\subsection{CCSNe: Surrounding Ejecta} \label{ccsne}

\citealt{LAMB2004} proposed that both GRBs and X-ray flashes (XRFs) stem from narrowly collimated jets produced by the collapse of massive stars, implying a population of events similar to FXTs. Early evidence supporting this idea has emerged from follow-up studies of \textit{EP} FXTs \citep{2025vandalen, 2025rasti, 2025eylesferris}. However, some low-redshift EP FXTs do not show signs of a SN (e.g., Rayson et al., \textit{in prep}) which makes it unlikely that all FXTs are the result of collapsars. While the collapse of massive stars is indeed a potential formation channel for FXT-producing millisecond magnetars, this scenario faces significant observational challenges, which we discuss in detail below.

CCSNe typically eject 1-15~$\rm{M_\odot}$ of material, expanding at velocities of $\sim 10^4~\rm{km~s^{-1}}$ \citep{Haynie_2023, Zha_2024}, resulting in an optically thick envelope that remains opaque to X-rays for weeks to months, thus preventing the observation of the characteristic X-ray emission of FXTs. As the SN ejecta expands and becomes less dense, the X-rays are absorbed by the surrounding ejecta and thermalized into optical radiation \citep{metzger2015_ccsne}. This thermalization mechanism is well-established in the context of superluminous supernovae (SLSNe), for which millisecond magnetars are considered to be the central engines powering them by converting the initial X-ray emission into optical radiation. This connection between millisecond magnetars and SLSNe is supported by observations such as SLSNe SCP06F6 \citep{Levan_2013}, which showed an X-ray outburst with a peak luminosity of $\rm L_X \sim 10^{45}~\rm{erg~s^{-1}}$ and a duration of $\lesssim~10$~ks, similar to the expected FXT luminosities and durations. 

The general yet uncertain idea is that the conservation of angular momentum during the collapse of the massive star causes the core to spin faster, which creates a jet that emerges from the collapsing star. It is collimated and accelerated by the magnetic field, and it propagates through the surrounding SN ejecta (\citealt{2016levan}; and references therein). This could provide a pathway for the X-ray emission to escape the dense SN ejecta, while simultaneously interacting with the surrounding medium to produce optical emission \citep{Margalit_2018}. For example, recent observations of EP240414a demonstrate that jet formation within a dense stellar envelope can produce X-ray outbursts reaching luminosities of $\rm \sim 10^{48}~erg~s^{-1}$ \citep{vandalen2024einsteinprobetransientep240414a}. However, previous studies have raised important challenges to the rapid-rotation scenario for magnetar formation in CCSNe. Simulations suggest that magnetic fields can be amplified to magnetar field strengths even in slowly-rotating progenitors through small-scale dynamo action \citep{2020muller_varma}. Moreover, observational studies of SN remnants containing magnetars show no evidence of enhanced explosion energies that would be expected from rapidly rotating NSs, or millisecond magnetars. These remnants show explosion energies consistent with $\rm \sim 10^{51}~erg$, suggested initial spin periods $>5$~ms \citep{2008vink}. While fallback accretion could potentially mask some signatures of rapid rotation by modifying the explosion energetics and remnant properties (e.g., \citealt{Metzger_2018, Wei_2021}), the combined theoretical and observational constraints suggest that extreme rapid rotation may not be necessary for magnetar formation \citep{2016torres_forne}. 

Another key factor to consider is the metallicity of the star, which plays a critical role in both the formation of the central engine and the jet collimation (if it is indeed produced). Lower metallicity stars tend to retain more of their initial angular momentum due to reduced mass loss through stellar winds \citep{2011georgy} which leads to rapid rotation, and are more likely to collapse into millisecond magnetars \citep{2016levan} (see their figure 7). Consequently, the formation efficiency of millisecond magnetars may exhibit a metallicity dependence, with potentially higher rates occurring in low-metallicity stars \citep{Song_2023}. Further, in low-metallicity stars, the jet is probably more narrowly collimated \citep{Mizuta_2009}. Hence, the massive stars that collapse to produce millisecond magnetars tend to generate highly collimated jets reducing the likelihood of detecting FXTs from these events, and may also explain the non-detection of gamma rays in some FXTs.

\subsection{Implications for the Origin of FXTs}
To consider the implications for the origin of FXTs, it is important to first note that not all FXTs are alike and they potentially arise from different origins, for example, XRT 110621 is suggested to be from a SN SBO origin \citep{alplarsson2020, deepak2024}, XRT 100831 is suggested to be a WD-IMBH TDE \citep{quirola1, quirola_bns, inkenhaag2024redshiftscandidatehostgalaxies} and XRT 030206 possibly has a BNS origin \citep{alplarsson2020, deepak2024}; this implies that only a fraction of all FXTs can be linked to a millisecond magnetar origin. 

Some detected FXTs were found to be associated with GRBs, as confirmed by contemporaneous GRB detections, for example, EP240315a \citep{2024levan} and EP240219a \citep{2024yin}. Population synthesis studies of Galactic magnetars reveal that $\rm \leq 16~Myr^{-1}$ (millisecond) magnetars could have formed via GRB-like events \citep{Rea_2015}. Using the method described in Section~\ref{rateconversion}, this corresponds to, and provides us with an upper limit of $\rm \mathcal{R}_{SFR} \approx 10^2 - 10^3~Gpc^{-3}~yr^{-1}$ and $\rm \mathcal{R}_{SMD} \approx 10^1 - 10^2~Gpc^{-3}~yr^{-1}$. Given that the detected FXT rate is $\rm \approx 10^3 - 10^4~Gpc^{-3}~yr^{-1}$\footnote{\cite{quirola1, quirola2} calculate the FXT event rate up to $\rm z \sim 2$, whereas we are extrapolating $\rm \mathcal{R}_{SFR}$ and $\rm \mathcal{R}_{SMD}$ to $\rm z \sim 4$, however this does not alter the conclusions we present here.}, this constraint suggests that millisecond magnetars can account for at most 10\% of the total detected FXT population. 

Further, we compare the Galactic magnetar rate of $\rm \leq 16~Myr^{-1}$ to the rate of formation of millisecond magnetars from BWD mergers in the MW, i.e., $\sim$300~$\rm Myr^{-1}$ \citep{levan2006}, and the BNS merger rate in the MW, $\sim$30~$\rm Myr^{-1}$ \citep{sgalletta2023}. The lower rate of Galactic magnetars implies that most BWD and BNS merger products must either collapse promptly to form BHs or form unstable magnetars that quickly collapse. Finally, the $\rm \sim 20~Myr^{-1}$ rate from the collapse of massive stars (Section~\ref{massivestars}) aligns with the Galactic magnetar rate, positioning this channel as a plausible, although rare, contributor to the FXT population.

\subsection{Implications for the Origin of Other Transients}
Millisecond magnetars have been proposed as the central engine for several classes of energetic transients beyond FXTs (e.g., \citealt{metzger2015_ccsne}). These include superluminous supernovae (SLSNe) (e.g., \citealt{2015nicholl_slsne, 2021vurm_slsne, 2024gottlieb_slsne}), long and short gamma-ray bursts (LGRBs and SGRBs) (e.g., \citealt{2015_msec_sgrbs, 2016levan, 2020sarin}), fast radio bursts (FRBs) (e.g., \citealt{2017metzger_frb, Nicholl_2017, 2020frbhosts_magnetars}), and fast blue optical transients (FBOTs) (e.g., \citealt{2022yao_fbots, 2024long}). The rapid rotation and strong magnetic fields of millisecond magnetars can power these events through different mechanisms, e.g., the spin-down energy injection can produce the extraordinary luminosity of SLSNe, or drive the relativistic outflows in GRBs, and generate coherent radio emission in FRBs. Each transient class occupies a distinct region in the parameter space of magnetic field strength, initial spin period, and ejecta mass, suggesting different formation channels. While our derived constraints on the millisecond magnetar formation rates could have important implications for the rates of these transient populations, a detailed comparison of their event rates and the required magnetar formation efficiency for each class extends beyond the scope of this work.

\section{Conclusion} \label{conclusion}

FXTs are short X-ray flashes with durations of a few minutes to hours, that have been detected by telescopes such as \textit{Chandra, XMM-Newton}, and \textit{Swift-XRT} at a rate of $10^3 - 10^4~\rm{Gpc^{-3}~yr^{-1}}$ over the last two decades. With the launch of \textit{Einstein Probe (EP)} in early 2024, there have been many more FXT detections even during its commissioning phase (that ended in July 2024) and since the start of nominal operations. While several theories have been proposed, the origin of FXTs is not yet understood. One model invokes the spin-down of a highly magnetic, millisecond spin NS, or a millisecond magnetar, produced as a post-merger remnant of a BNS merger. In this paper, we consider all possible formation pathways for millisecond magnetars that can produce an FXT, such as the AIC of massive WDs, BWD, NSWD and BNS mergers, and the collapse of massive stars, and compare it to the FXT rate. We find that the highest rate of formation of (millisecond) magnetars is from massive stars, followed by BWD mergers and BNS mergers. Several key factors limit the viability of millisecond magnetars as FXT progenitors such as the uncertainties in the spin and magnetic field distributions of magnetars in all the scenarios we have discussed, and the presence of dense surrounding ejecta preventing the detection of X-rays on timescales compatible with the duration of FXTs along with other observational constraints, in the case of CCSNe. The requirement for both rapid rotation on $\sim$~millisecond timescales, and strong magnetic field strengths of $10^{14} - 10^{15}$~G, significantly reduces the fraction of events that could result in the formation of millisecond magnetars, and in several cases, the exact fraction is difficult to precisely constrain. With \textit{EP} now detecting several FXTs, the diversity in the FXT properties suggests that FXTs arise from different origins, and millisecond magnetars can account for at most 10\% of the entire FXT population. Future observations, particularly with rapid multi-wavelength follow-up, and the identification of the host galaxies of FXTs, will be crucial for determining the relative contributions of different progenitor scenarios.

\begin{acknowledgements}
SB would like to thank Ilya Mandel for his helpful inputs during discussions related to this project. SB acknowledges studentship support from the Dutch Research Council (NWO) under the project number 680.92.18.02. PGJ is supported by the European Union (ERC, StarStruck, 101095973). Views and opinions expressed are however those of the author(s) only and do not necessarily reflect those of the European Union or the European Research Council. Neither the European Union nor the granting authority can be held responsible for them. This work made use of Python packages \texttt{NUMPY} \citep{numpy}, \texttt{SCIPY} \citep{scipy}, and \texttt{MATPLOTLIB} \citep{matplotlib}. This work made use of \texttt{ASTROPY}: a community-developed core Python package and an ecosystem of tools and resources for astronomy \citep{astropy:2013, astropy:2018, astropy:2022}. We thank the referee for the insightful comments on this manuscript.
\end{acknowledgements}

%
%

\bibliographystyle{aa}
\bibliography{bibl}





\end{document}